  \documentclass{book}
  \usepackage[utf8]{inputenc}
  \usepackage[T1]{fontenc}
\usepackage{amsmath}
\usepackage{revsymb}
\usepackage{amssymb} 
\usepackage{amsfonts}
\usepackage{frenchle}
\usepackage{fancyhdr}
\usepackage{graphicx}
\usepackage{makeidx}
\usepackage[labelformat=simple,labelfont=it]{subfig}
\usepackage{yhmath}
\usepackage{textcase}	
\usepackage{float}
\usepackage{calc}
\usepackage{framed}
\usepackage{xcolor}
\usepackage{pst-grad}
\usepackage{pst-slpe} 
\usepackage{pst-ob3d}      
\usepackage{url}

\usepackage{bm} 

\input{macros2.sty}

\def\dd{{\rm d}}

\def\dd{{\rm d}} 

\begin{document}

\bigskip
\bigskip

\centerline{\bf \large Phase velocity and light bending in a
  gravitational potential} \bigskip

\centerline{Jos\'e-Philippe P\'erez and Brahim Lamine} 
\medskip

Institut de Recherche en Astrophysique et Planétologie (IRAP)-CNRS-Universit\'e de 
Toulouse- 14 avenue \'Edouard Belin, 31400 Toulouse, France
\smallskip

jose-philippe.perez@irap.omp.eu $\quad\quad$
brahim.lamine@irap.omp.eu
 
\bigskip
\bigskip

{\bf Abstract}

In this paper we review the derivation of light bending obtained
before the discovery of General Relativity (GR). It is intended for
students learning GR or specialist that will find new lights and
connexions on these historic derivations. Since 1915, it is well known
that the observed light bending stems from two contributions~: the
first one is directly deduced from the equivalence principle alone and
was obtained by Einstein in 1911; the second one comes from the
spatial curvature of spacetime. In GR, those two components are equal,
but other relativistic theories of gravitation can give different
values to those contributions. In this paper, we give a simple
explanation, based on the wave-particle picture of why the first term,
which relies on the equivalence principle, is identical to the one
obtained by a purely Newtonian analysis. In this context of wave
analysis, we emphasize that the dependency of the velocity of light
with the gravitational potential, as deduced by Einstein concerns the
phase velocity. Then, we wonder whether Einstein could have envisaged
already in 1911 the second contribution, and therefore the correct
result. We argue that considering a length contraction in the radial
direction, along with the time dilation implied by the equivalence
principle, could have led Einstein to the correct result. \bigskip
\bigskip

\section*{Introduction}
\medskip

The Newtonian theory of the deviation of light bending was published
in 1801 by the German physicist
J. Soldner~\cite{Soldner1801,Hadley1966}. The author develops Kepler's
classical motion of a particle of light, of mass $ m $, submitted to
the gravitational force exerted by a mass $M$ with spherical
symmetry. He obtained the usual hyperbolic motion and computed the
deflection angle $ \chi_N $ of the trajectory in the Newtonian
approximation. By applying this analysis to a particle of light
grazing the Sun, he found the value $ \chi_N \approx 0.87 \ut {as} $,
which is exactly half of the experimental value measured in 1919
\cite{Dyson1919}. In the first section, we review the computation of
Soldner, with modern notations.

In 1911~\cite{Einstein1911}, Einstein proposed a new analysis of light
bending, based on the equivalence principle alone. He was led to the
conclusion that a dilation of duration is produced by a gravitational
potential. This leads to the conclusion that a certain velocity of
light should depend on the gravitational potential~$\Phi$,
\begin{equation}
\label {eq:0}
c_{p, \Phi} = c \left (1 + {\Phi \over c^2} \right)
\end{equation}
This velocity is smaller than $ c $, the value in the absence of
potential ($ \Phi = - GM /r < 0 $). In his original paper of
1911~\cite{Einstein1911}, Einstein does not give a real physical
interpretation of this velocity, but simply speaks of \og speed of
light\fg{}. Using the principle of Huygens-Fresnel, he deduced the
trajectory of a light ray by requiring that they are normal to wave
front. Curiously, he found the same expression as the Newtonian result
of Soldner. In the second section, we review the Einstein argument in
a slightly different way, which shed new light on the Einstein
derivation. In particular, we show that the velocity obtained by
Einstein has to be interpreted as a phase velocity, and not the light
speed (that remain a fundamental constant). We then argue that the de
Broglie wave transposition of Soldner's analysis explain the identical
result obtained by Einstein.

Only a few years later, as part of the complete theory of general
relativity~\cite{Einstein1915,Einstein1916}, Einstein obtained the
correct value of this deviation, i.e. the double of the previous
result. Many authors have discussed the reason of the doubling of the
Newtonian result in GR~\cite{Lerner1997}. In the third part of this
paper, we propose a new light to interpret this doubling. For this, we
propose a generalization of the physical analysis of Einstein,
accompanying the time dilation due to a gravitational potential, by a
concomitant contraction of the radial lengths (see~\cite{Provost2016}
where this idea has already been proposed, though with a different
approach as we do). This derivation is an intuition that could have
had Einstein, more than a formal proof, because it is already known
that the correct result cannot be recovered simply from the
equivalence principle and the Newton's limit alone~\cite{Gruber1988}.

\bigskip

\section{Newton theory of Soldner}

In this section, we briefly summarize how Soldner computed light
bending by a massive body from a Newtonian approach. For a complete
historical perspective about the Newtonian influence of gravitation on
light, see~\cite{Eisenstaedt1991}. For this, he
hypothetized that light is made of material particles, for which it is
possible to apply Newton's laws in order to obtain the trajectory. To
justify his hypothesis, he added, in the part related to the
objections which might be opposed to him, that light should be
considered as matter~:

{\it \og Hopefully, no one would find it objectionable that I treat
  a light ray as a heavy body. That light rays have all the absolute
  [basic] properties of matter one can see from the phenomenon of
  aberration which is possible only because light rays are truly
  material. And furthermore, one cannot think of a thing which exists
  and works on our senses that would not have the property of
  matter.\fg{}}

The computation of Soldner is prior to Maxwell's theory, in which the
speed of light is a constant~\footnote{In particular, it is
  independent of the gravitational potential, because gravitation and
  electromagnetism are not coupled in this theory.}. In Soldner's
perspective, the speed of a particle of light is not a constant, but
varies along the path around the massive body, just like an ordinary
material particle. In his publication, there is therefore a free
parameter, which he took as being the speed of light measured at the
level of perihelion $P$; in the following, we will note this velocity
as $ v_P $.

\begin{figure}\centering
\includegraphics [width = 10cm]{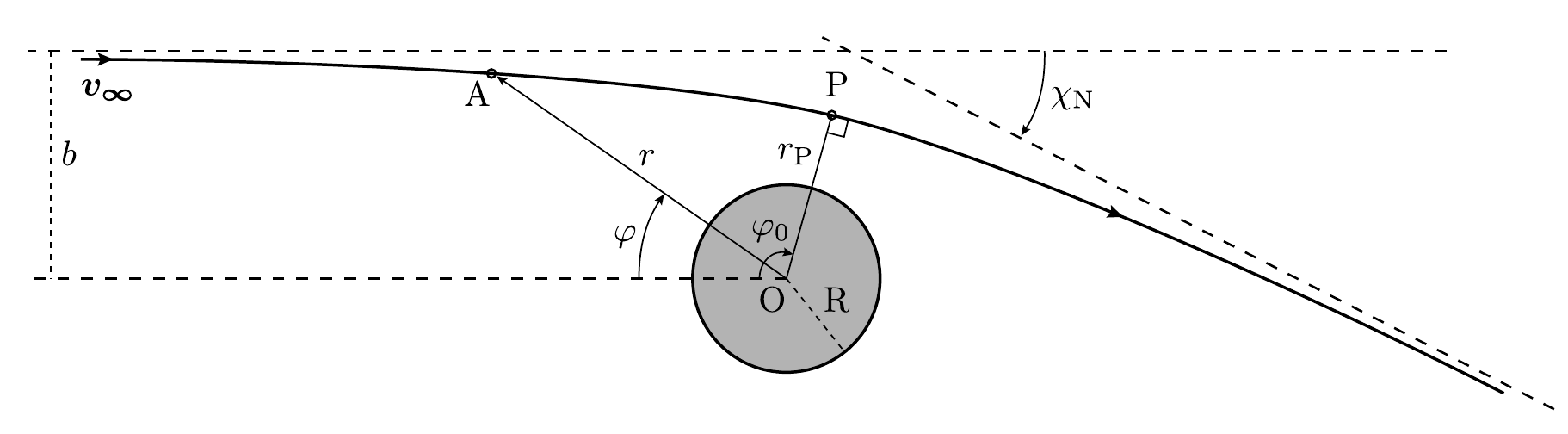}
\caption{Diagram  of deviation of a light ray by a mass with spherical
  symmetry. Notations are defined in the text.}
\label{cg-soldn}
\end{figure}

The trajectory of a particle of light $ A $ can be deduced from the
conservation of the massic mechanical energy
$e_m={v^2 / 2} - {GM / r}$ and the massic angular
momentum $ \vg{\ell}=r^2 \dot \varphi \, \vg {e}_z$. In these
expressions, $ r $ and $ \varphi $ are the polar coordinates of $ A $,
in the plane of motion defined by $O$ and the normal vector
$ \vg{\ell} $ (Fig. \ref{cg-soldn}). Combining these two expressions give~:
\begin{equation}
\label{eq:cons E}
\dot r^2 + {\ell^ 2 \over r^2} - 2 {G M \over r} = 2 e_m
\end {equation}

The mass $ m $ of the particle of light, which was unknown to Soldner
in 1801, does not appear in this equation. This observation is simply
a reformulation, in the case of light, of the underlying hypothesis of
the equality of the gravitational mass, which appears in gravitational
energy, and of the inertial mass, present in the angular
momentum. This hypothesis was early postulated by Galileo and then
tested experimentally, with a relative precision of $10^{-3}$, by
Newton using pendulums made of different materials.

Following Soldner, the constant $ \ell $ can be expressed with respect
to the speed of light $ v_P $ at perihelion,
$ v_P = r_P \dot {\varphi} $. Likewise, one can also introduce the
impact parameter $ b $, so that $ \ell = r_P v_P = b \, v_\infty $ (see
Fig.~\ref{cg-soldn} for notation). Equation~(\ref{eq:cons E}) can be
rewritten using dimensionless quantities. Introducing
$ \rho \equiv r / r_P $ (as Soldner did), expression~(\ref{eq:cons E})
is written more conveniently, if we introduce the gravitational
potential at perihelion $ \Phi_P \equiv -GM / r_P $, as
\begin{equation}
\label{eq:40}
r_P^2 \left (\frac {\dd \rho} {v_P \dd t} \right)^ 2 +
\frac {1} {\rho^2} + \frac {2} {\rho} \frac {\Phi_P} {v_P^2} =
\frac {2e_m} {v_P^2} = \left (\frac {v_\infty} {v_P} \right)^2
\end{equation}
This equation is identical to the one obtained by Soldner. He solved
the equation~(\ref{eq:40}) with lenghty calculations, because the
usual Binet change of variable was not yet known. Using the reduced
Binet variable $ u \equiv 1 / \rho = r_P / r $, one finds
from~(\ref{eq:40})~:
\begin{equation}
\label{eq:40bis}
{\dd^2 u \over \dd \varphi^2} + u = - \frac {\Phi_P} {v_P^2} \equiv
{r_P \over p}
\end{equation}

If the light is grazing on the surface of the attractive body,
$ r_P = R $, with $R$ the radius of the massive body. The
dimensionless quantity $ - \Phi_P / v_P^2 $ is positive and reduces to
the compactness $ \mathcal {C} \equiv GM / (R v_P^2) $ of the object,
which physically represent the ratio between the gravitational energy
and the mass energy. For objects like planets or stars, the
compactness is very small compared to unity, so that the right-hand
side of equation~(\ref{eq:40bis}) is very small and the solution is
nearly the usual Newton solution. For the Sun and the Earth, we find
respectively (taking $ v_P \approx c $)~:
\begin{equation}
\label{eq:30}
\mathcal {C}_{\odot} \approx 2 \times 10^{- 6}
\quad\quad\hbox{and}\quad\quad
\mathcal {C}_{\oplus} \approx 7 \times 10^ {- 10}
\end{equation}

The solution of equation~(\ref{eq:40bis}) is given by
$u (\varphi) = A \cos (\varphi- \varphi_0) + {r_P / p}$.  We then
determine the constant $ A $ using the condition on perihelion,
$ u = 1 $ when $ \varphi = \varphi_0 $, which gives
$ 1 = A - \Phi_P / v_P^ 2 $. Thus, the solution for $ r $ is a conic
of parameter $ p $ and eccentricity~$e$~:
\begin{equation}\label{eq:ecc}
r = {p \over 1 + e \, \cos (\varphi - \varphi_0)}
\quad \text {with}\quad
p = r_P \, \left(\frac {v_P ^ 2} {- \Phi_P} \right)
\quad \text{and}\quad
e = \left(\frac {v_P ^ 2} {- \Phi_P} \right) - 1
\end{equation}
We can also relate $ e $ with the massic mechanical
energy ~:
\begin{equation}
e_m = \frac{v_P^2} {2} \left (1 + \frac {2 \Phi_P} {v_P^2} \right) =
\frac{v_P^2} {2} \left ({e - 1 \over e + 1} \right) =
\frac {v_P^2} {2} \left (\frac {\Phi_P} {v_P^2} \right)^2 (e^2 - 1)
\end{equation}
The previous expression allows to study the type of trajectories as a
function of the value of the eccentricity~: hyperbolic motion for
$e_m>0$ ($e>1$) parabolic motion for $e_m=0$ ($e=1$) and elliptic
motion for $e_m<0$ ($e<1$). Soldner found that in pratice $ e_m> 0 $,
because the condition $ - \Phi_P / v_P^2 \ll 1 $ was satisfied for the
stars known at that time. Therefore $e\gg1$ according to
eq. (\ref{eq:ecc}) and the corresponding trajectories of the particles
of light are hyperbolic ones. 

Soldner briefly evoked the existence of bounded solutions,
characterized by $ e_m < 0$, i.e.  $GM / (r_P v_P^2)> 1/2$.  He added,
however, that this condition was not realistic, or in any case it did
not correspond to any known object at that time~\footnote{Soldner
  wrote {\it \og Since it does not matter how much mass it would be so
    great that it could produce such an acceleration gravity, a light
    ray describes, in the world known to us, always hyperbola. \fg{}}
  We will discover much later that such objects, for which the
  trajectory of light realizes $ e_m <0 $, do exist in nature, for
  example black holes. Note that Michell already considered bounded
  trajectory of light, but in a rather different situation : he
  considered radial trajectory of light from massive objects,
  from which the escape velocity would be greater than the speed of
  light~\cite{Eisenstaedt1991}.}. Indeed, the stars seen in the sky
were already considered as sun-like, whose mass and radius were known
with sufficient precision. The compactness should be of the same order
of magnitude than $\mathcal{C}_\odot$, and therefore very small (see
equation (\ref{eq:30})).

The Newtonian deviation angle $ \chi_N $ is easily obtained by writing
the asymptotic condition $ r \rightarrow \infty $, i.e.
$ \cos (\varphi_{\text{in}}- \varphi_0) = -1 / e $. By choosing
$ \varphi_{\text{in}} = 0 $ for the direction of the incident ray, the
ray emerges asymptotically in $ \varphi = \pi + \chi_N $, so that
$\cos \varphi_0 = \cos (\pi + \chi_N - \varphi_0) = - 1/e$. Hence
$ \chi_N = 2 \varphi_0 - \pi $ and therefore
$\tan \varphi_0 = \tan \left ({\chi_N / 2} + \pi/2\right) = -
\tan^{-1} (\chi_N / 2)$. Since $ \cos \varphi_0 = -1 / e $,
$ \tan \varphi_0 = - (e^2 - 1)^{1/2} $, and we find the following
result of Soldner~:
\begin{equation}
\label{eq:11}
\tan \left (\frac {\chi_N} {2} \right) = \frac {1} {(e^2 - 1)^{1/2}} =
\frac {- \Phi_P / v_P^2} {(1 + 2 \Phi_P / v_P^2)^{1/2}}
\end{equation}
This Newtonian result is an exact result, which does not rely on any
assumption. In the limit $ - \Phi_P / v_P^2 \ll 1 $, it gives
$\chi_N \approx 2GM/(r_P v_P^2)$. Or, since
$ r_P \, v_P = b \, v_\infty $ and $ r_P \approx b $ (at lowest order
in $ - \Phi_P / v_P^2 $)~:
\begin {equation}
\chi_N \approx \frac {2GM} {b \, v_\infty^2} =
{r_S \over b} \left ({c \over v_{\infty}} \right)^2
\quad\hbox{where}\quad
r_S = {2 GM \over c^2}
\end {equation}
is the Schwarzchild radius. In order to estimate the orders of
magnitude, Soldner used the speed of light measured by Bradley in
1729, using the aberration of stars~\cite{Bradley1729}~\footnote{Note
  that Bradley obtained this speed, in unit of speed of the Earth
  around the Sun, the latter being poorly known at the time.}. The
result obtained by Soldner is half the one predicted by general
relativity in 1915 \cite{Einstein1915}. Moreover, its expression
(\ref{eq:11}) is not universal, because it involves the speed of light
at perihelion (or equivalently $ v_\infty $), the latter being not
considered, at the time of Soldner, as a universal constant. However,
Soldner seems to suppose that this speed, which is much greater than
the speed of celestial objects (planets, stars), must be, according to
the law of Galilean composition of velocities, quite close to the
value which he used in its numerical applications (see also the
discussion in~\cite{Eisenstaedt1991} who takes up the argument of
Michell about the variation of the speed of light in a gravitation
field). Assuming that $ v_{\infty} \approx c $, one obtains, if the
light is grazing, for the Sun and the Earth respectively ~:
$$
\chi_{N, \odot} \approx 0.87 \ut {as}
\quad\hbox{and}\quad
\chi_{N, \oplus} \approx 0.28 \times 10^{- 3} \ut {as}
$$
Soldner deduced from these numerical results that the deviation of
light near the Sun was too small to be measured at his
time~\footnote{He concludes with this sentence~: {\it \og So it is clear
    that nothing is necessary, at least in the present state of
    practical astronomy, that one should take into account the
    disturbance of light rays by attracting celestial bodies.\fg{}}}. He
(unknowingly) announced a result that will be tested experimentally
more than a century later \cite{Dyson1919}. It is interesting to note that he
publishes the result of his analysis, even if the conclusion of this
one is that the effect is not observable~\footnote{He even adds in its
conclusion: {\it \og At any rate, I do not believe that there is any
  need on my part to apologize for having published the present essay
  just because the result is that all perturbations are unobservable.\fg{}} }.

\section{Einstein relativistic theory of 1911}

Einstein already noticed in 1907, in his review article on special
relativity, that, according to the principle of equivalence, a light
ray has to be bent by gravitation~\cite{Pais1993}. In 1911 he
carefully studied the influence of a gravitational potential $ \Phi $
on the propagation of light in vacuum. For a review of the original
derivation, see~\cite{Darrigol2015}. He based its arguments on two
pillars~:
\begin{itemize}
\item special relativity, including Maxwell theory of
  electromagnetism. It contains in particular the universal character
  of the speed of light in vacuum and the Doppler-Fizeau effect.
\item the equivalence principle he developed to build the theory of
  general relativity; this principle affirms the equivalence between
  an observer at rest in a uniform gravitational field and an observer
  uniformly accelerated in the absence of gravitation
  (see~\cite{Brown2016} for philosophical considerations concerning the
  principle of equivalence).
\end{itemize}

\begin{figure}\centering
\includegraphics [width = 8cm] {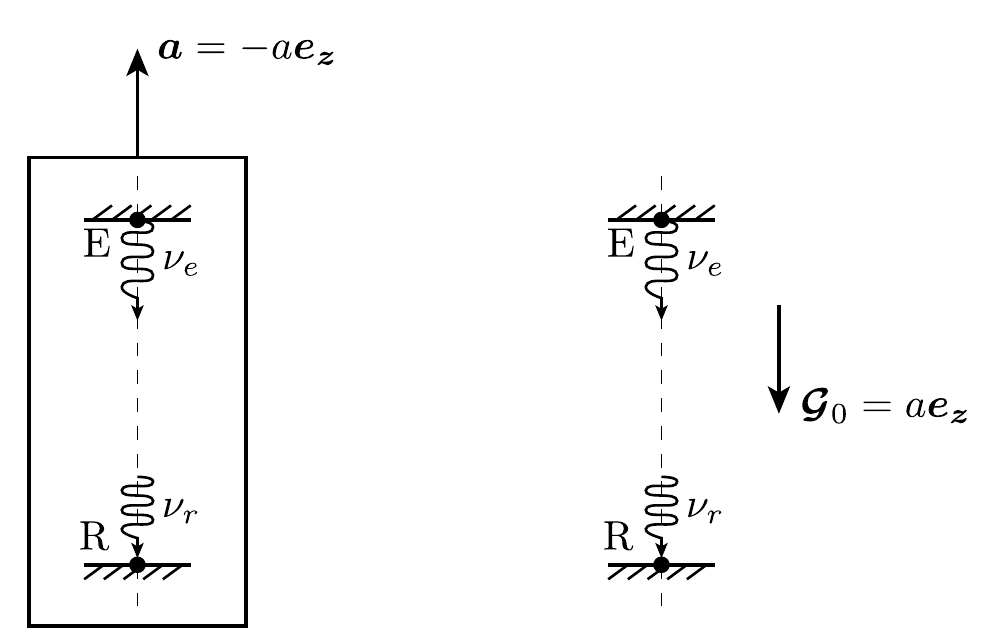}
\caption{Diagram of the experience of Pound and Rebka and illustration
  of the equivalence principle. Notations are defined in the text.}
\label{cg-pound}
\end{figure}

Inspired by Einstein's reasoning let us consider two observers, each
one having a clock of the same manufacture. These two observers are
assumed to have a uniform acceleration $ a $, for example by being
both in the same rocket subjected to this acceleration. These two
observers exchange photons, from the emitter $ E $ to the receiver
$ R $ located at a distance $ H $ (Fig. \ref{cg-pound} on the
left). Due to the Doppler-Fizeau effect, the frequency $ \nu_r $ of
the electromagnetic wave received by $ R $ differs from the frequency
$ \nu_e $ of the wave emitted by $ E $. At lowest order (ignoring
relativistic corrections which would produce a negligible second-order
effect here), the photon is received by $ R $ after a time interval
$ H/c $. The velocity of $ E $ is then $ v = a H/c
$. As a result, according to the Doppler-Fizeau effect, the relation
between $ \nu_r $ and $ \nu_e $ is (still at lowest order)~:
\begin{equation}
\nu_r = \nu_e \left (1 + \frac {v} {c} \right) =
\nu_e \left (1 + \frac {a H} {c^2} \right)
\end{equation}

Because of the equivalence principle, the situation in an accelerated
rocket is physically equivalent to the one of rest observers in a
uniform gravitational field $ \vg {\G} = \G_0 \, \vg{e}_z $, such that
$ \G_0 = a $ (Fig. \ref{cg-pound} on the right). We remind that $\G$
is such that the newtonian gravitational force $F$ exerted on a mass
$m$ submitted to the gravitational field is $F=m\G$. Introducing now
the gravitational potential $ \Phi $, one has
$\Phi_e - \Phi_r = \G_0 H > 0$. The gravitational potential is
related, up to a constant, to the gravitational potential energy of a
mass $m$ in the gravitational field by the relation $\E_p=m\Phi$. Thus ~:
\begin{equation}
\label{eq: freq}
\nu_r = \nu_e \left (1 + \frac {\Phi_e - \Phi_r} {c^2} \right)
\quad\hbox{or}\quad
\nu_r \left (1 + \frac {\Phi_r} {c^2} \right) =
\nu_e \left (1 + \frac {\Phi_e} {c^2} \right)
\end{equation}
to first order~\cite{Perez2016}.

This theoretical prediction of Einstein has been tested experimentally
for the first time by Pound and Rebka in 1960~\cite{Pound1959}. In his
article written in 1911, Einstein proposed to measure this effect
using the shift of the spectral lines of the Sun, while emphasizing
that the effect was very small since
$ \mathcal {C}_{\odot} \approx 2 \times 10^{- 6} $.

According to Einstein, equation~(\ref {eq: freq}) does not express
just a simple Doppler-Fizeau effect on an electromagnetic wave, but
more fundamentally an influence of the gravitational potential on
time. To reach this conclusion, one can argue that the number of
oscillation cycles in a wave packet exchanged between $E$ and $R$ must
be preserved~\footnote{Likewise, Einstein argued that the number of
  nodes and antinodes between $ E $ and $ R $, when a standing wave is
  established between the transmitter and the receiver, has to be
  constant, otherwise we would be in the presence of a non-stationary
  process, which is excluded.}. Therefore, introducing the proper
durations $ \tau_e $ and $ \tau_r $ measured by clocks in $ E $ and
$ R $, one has $\nu_r \, \dd \tau_r = \nu_e \, \dd \tau_e$, that is to
say $\nu_ \Phi \, \dd \tau_ \Phi = \text{Cte}$ or equivalently~:
\begin{equation}
\label{eq: tau phi}
\frac {\dd \tau_ \Phi} {1 + \Phi / c^2} = \dd \tau_0
\end{equation}
$ \tau_0 $ being the proper duration measured by a distant observer,
located at a point for which $ \Phi \approx 0 $ (typically at
infinity). What is true for the photon frequency must be true for all
other fields~: in other words it is the proper duration $ \tau_ \Phi $
that flows differently for $ E $ and for $ R $.

The dependency of $ \tau_{\Phi} $ with the gravitational potential has
of course to remain compatible with the foundations of the special
relativity and the equivalence principle. It implies, in particular,
that the speed of light, as measured by a observer at the point where
he stands (this precision is important), has to stay equal to $ c $,
\begin{equation}
  \label{eq: 13}
  \frac {\dd r} {\dd \tau_ \Phi} = c
\quad \hbox {which implies} \quad
\frac {1} {1+ \Phi / c^2} \, \frac {\dd r} {\dd \tau_0} = c
\end{equation}
Hence the speed of light $c_{p,\Phi}$ measured by a \emph{distant}
observer (with proper time $\tau_0$), who observes the propagation of
the latter in the vicinity of a massive star, will be~\footnote{Note
  that this relation, relativistic in essence, supposes that the
  gravitational potential $ \Phi $ is defined without additive
  constant; in Newtonian mechanics, the effect of the constant is
  neutralized by the infinite value of the speed of propagation of
  light.}
\begin{equation}
  \label {eq: 12}
  c_{p, \Phi} = \frac {\dd r} {\dd \tau_0} = 
	c \left (1 + \frac{\Phi} {c^2} \right)
\end{equation}
Einstein obtained this expression in 1911~\cite{Einstein1911} with a
different argument. Nevertheless, in his paper, he was not clear about
the physical interpretation of this velocity. In particular, he was a
little bit embarassed with the fact that special relativity and the
equivalence principle has to imply a constancy of the speed of light,
while its result shows in the contrary a dependency with the
gravitational potential. He even wrote that \go {the principle of the
  constancy of the speed of light is not valid in the sense that
  serves as a basis for the usual theory of relativity \gf}. In fact,
there is no inconsistency with special relativity and the key point
here is that this velocity $c_{p,\Phi}$ is relevant only for a distant
observer. An observer at the level of the perihelion would indeed
measure that the speed of light is equal to $c$ at this point, and
this is not in contradiction with equation~(\ref{eq: 12}). The second
key ingredient is that this velocity is in fact a \emph{phase}
velocity. Einstein did not mention this term in his paper of 1911,
where he used the generic term \og speed of light\fg{} without
distinguishing between phase or group velocity. If this velocity is
interpreted as a group velocity, it would imply that light would be
bend in the opposite direction, that is to say outwards instead of
towards the central body!

Hopefully, Einstein used a wave analysis of the bending, and therefore
arrived to a bending towards the central mass. To do this, he
considered the propagation of a wave front propagating at velocity
$c_{p,\Phi}$ in a non uniform gravitational potential, and deduced the
trajectory of light through the Malus theorem. We adopt here another
approach, based on the eikonal equation.

Indeed, the dependency of the phase velocity of light with a
gravitational potential $ \Phi $ can also be interpreted in terms of
an effective refraction index $ n_ \Phi $ of the (empty) medium in
which light propagates, according to~:
\begin{equation}
\label{eq:nphi}
n_ {\Phi} = {c \over c_{p, \Phi}} = {1 \over 1 + \Phi /c^2} \approx 
1 - {\Phi \over c^2} >  1
\end{equation}
In order to determine the trajectory, one can now use the eikonal
equation in the (approximation of the geometrical optics). Introducing
the Frenet base $ (\bm {e}_t, \, \bm {e}_n) $ and the curvilinear
abscisse $ s $ along the trajectory, the equation of the light ray
is given by~\cite{Perez:optique}~:
 \begin{equation}
\frac {\dd} {\dd s} \left(n_\Phi \, \bm {e}_t  \right) =
\grad n_\Phi
\end{equation}
Multiplying this equation by $ \bm {e}_n$ and introducing the
elementary deflection angle of the path,
$ \dd \chi = - \dd \bm {e}_n \cdot \vg {e}_t $, one gets~:
$$
n_ {\Phi} \, {\dd \chi \over \dd s} = 
- {1 \over c^2} \, \grad \Phi \cdot \bm {e}_n
$$

\begin{figure}\centering
\includegraphics [width = 8cm] {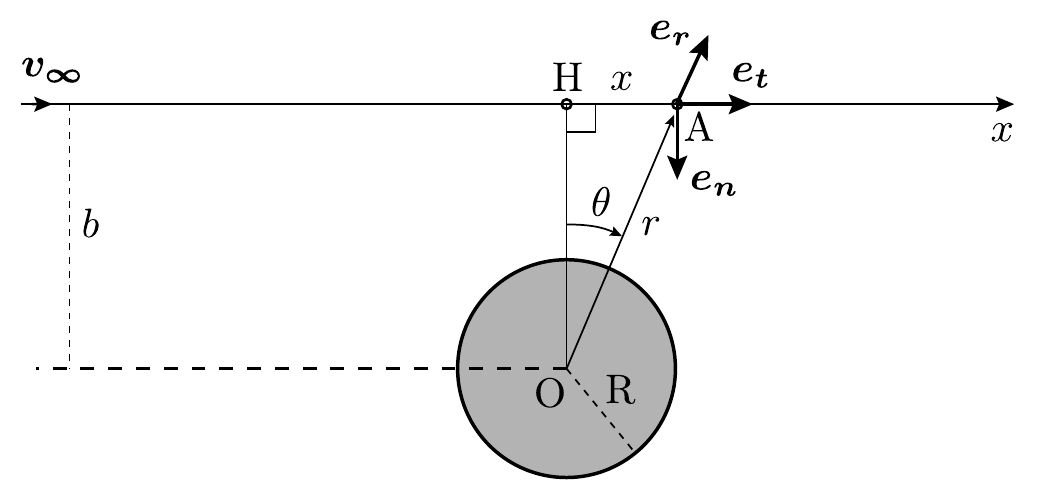}
\caption {Deflection of a light beam by a spherical mass distribution,
  according to Einstein in 1911. The straight line is the unperturbed
trajectory.}
\label{cg-dev11}
\end{figure}

Since $ \dd \chi / \dd s $ is of order $ 1 $, we can take
$ n_ {\Phi} \approx 1 $ at zeroth order. We then find the integral
expression of the deflection angle $ \chi_{E, 11} $ obtained by
Einstein in 1911~:
\begin{equation}\label{eq:integral}
\chi_{E, 11} \approx - \frac {1} {c^2} \int \grad \Phi \cdot \bm {e}_n \,
\dd s
\end{equation}
The minus sign indicates a deviation towards the massive
object. Treating $ \grad \Phi $ as a small perturbation, the previous
integral~(\ref{eq:integral}) can be computed on a straight line rather
than the actual curved trajectory. If we denote by $ x $ the
coordinate of the current point $ A $ on the trajectory, one gets,
using $ \Phi = - GM / r $ (Fig. \ref{cg-dev11})~:
\begin{equation}
\chi_ {E, 11} \approx \frac {GM} {c^2} \, \int {\cos \theta \over r^2} \, \dd x
\end{equation}
where $ \vg {e}_r \cdot \vg {e}_n = - \cos \theta $; the angle
$ \theta $ varies from $ - \pi / 2 $ to $ \pi / 2 $ when $ x $ moves
between $ - \infty $ to $ \infty $. Since $ r = b / \cos \theta $ and
$ x = b \tan \theta $, with $ b $ the impact parameter,
$ \dd x = b \, \dd \theta / \cos^2 \theta $ and therefore~:
\begin{equation}
\label{eq: chi E 11}
\chi_ {E, 11} = \frac {GM} {bc^2} \int_ {-\pi/2}^{\pi/2} \cos \theta \, 
\dd \theta={r_S \over b}
\end{equation}

This result is identical to the one of Soldner, although the
approaches adopted are substantially different. To understand the
reason, let us use the wave aspect of any physical object, based on
the Hamilton-Jacobi formalism and the link between the action $S$
associated to a particle and the phase $ \varphi = S / \hbar $ of the
associated wave~\cite{Perez:quantique}. The velocity of the particle
of light is given by $v^2 = v^2_ {\infty} - 2 \Phi$, so that
\begin{equation}
v = v_ {\infty} \left(1 - {2 \Phi \over v^2_\infty} \right)^ 
{1/2} \approx
v_ {\infty} \left (1 - {\Phi \over v^2_ {\infty}} \right) >  
v_ {\infty}
\end{equation}
As already mentioned, should this velocity be interpreted as a phase
velocity, it would give an effective refractive index $n_\Phi<1$ (see
equation (\ref{eq:nphi})), and therefore an opposite light bending
compared to observations. In order to determine the phase velocity
$ c_ {p, \Phi} $, we can consider the displacement of a wavefront
($\varphi=\text{cte}$) between $ t $ and $ t + \dd t $.
\begin{equation}
  \label{eq: 2}
  0=\frac {\dd \varphi} {\dd t} = \frac {\partial \varphi} {\partial t} +
\bm {c_ {p, \Phi}} \cdot \grad \varphi
\end{equation}
The displacement being perpendicular to the wavefront,
$ \bm {c_{p, \Phi}} $ is colinear to $ \grad \varphi $. Finally,
replacing $ \varphi $ with $ S / \hbar $, we deduce~:
\begin{equation}
  \label{eq: 3}
\frac {\partial S} {\partial t} + c_{p, \Phi} \left |\grad S \right | = 0
\end{equation}
This equation is analogous to the Hamilton-Jacobi
equation~\cite{Landau1969}, provided that $ c_{p, \Phi} $ is expressed
as a function of the generalized momentum. Then, one can use the fact
that the time derivative of the action is equal to the opposite of the
Hamiltonian, $ \partial S / \partial t = -H $. And because the
Hamiltonian does not depend explicitly on time, it is a constant
$ H = \mathcal {E} $ so that~:
\begin{equation}
 \label{eq:4}
c_ {p, \Phi} = \frac {\mathcal {E}} {\left |\grad S \right |} =
{\E \over p}
\end{equation}
where the generalized momentum $ p = \gamma m v$~\cite{Perez:relat} is
identified with $ \grad S $ in Hamilton-Jacobi formalism
($p_i=\partial S/\partial q_i$~\cite{Landau1969}). Combining the
previous equations gives finally~\footnote{The velocity $ v_{\Phi} $
  can be interpreted as the group velocity of the electromagnetic
  light wave, which allows to recover the well-known relation (\ref
  {eq:6}) on the product between the group and phase velocity.}~:
\begin{equation}
 \label{eq:6}
 v \times c_ {p, \Phi} = \frac {\mathcal {E}} {\gamma m} \approx c^2
\end{equation}
since $ \mathcal {E} = \gamma mc^2 + m \Phi \approx \gamma mc^2 $. It
can be seen that the mass of the particle disappears and that this
last relation is also valid for relativistic particles. It leads to
the following relation between the phase velocity in the presence of a
gravitational potential, and the phase velocity in its absence:
\begin{equation}
c_{p, \Phi} \approx {c^2 \over v_ {\Phi}} \approx 
{c^2 \over v_{\infty} (1 - \Phi / v^2_ {\infty})}\approx
c \left (1 + {\Phi \over c^2} \right)  < c
\end{equation}
where we used $\Phi/v_\infty^2\ll1$ and $ v_{\infty} \approx c$. This
expression of the phase velocity is exactly the same as the one
obtained by Einstein in 1911. As shown above, the wave associated to
the particle of light is the fundamental ingredient to understand the
identical results obtained by Soldner and Einstein. Note nevetheless
that the Einstein result is more universal because the speed of light
$c$ is a real constant of nature ($v_\infty$ is not).

\section {Einstein relativistic theory of 1915}

In 1915, Einstein re-analyzed, in the framework of his theory of
general relativity, the deviation of a light ray by a mass
distribution with spherical symmetry. He obtained a result which is
the double of what he initially published in 1911. In this new result,
a first contribution is attributed to the influence of the
gravitational potential on time (it is exactely the effect computed in
1911), and a second contribution, of the same magnitude, is related to
the deformation of space (spatial curvature). As already mentioned,
this new result was confirmed experimentally in 1919~\cite{Dyson1919}.

In this last section, we wonder whether Einstein could have come to
the right answer already in 1911. We first explain why the formal
answer is no, and then propose a guess that could have lead Einstein
to the track of general relativity before 1915.

To begin with, Einstein could not have established rigorously the
correct expression until he had completed the theory of general
relativity. The reason is that there are several possible relativistic
theories of gravitation, which are all in agreement with the
equivalence principle (see~\cite{Norton1992} for a review), but differ
from Einstein's GR. Also, different attempts have been made to simply
recover the Schwarzschild metric from the equivalence principle and
the Newtonian limit alone, but none succeeded~\cite{Gruber1988}. Only
experiments finally made it possible to decide in favor of Einstein
theory. All these relativistic theories of gravitation predict a first
contribution identical to the one obtained by the Newton approach (cf
equation (\ref{eq: chi E 11})). In GR, as already shown, this
contribution is understood as stemming from a curvature of time. The
difference lies in the second contribution, which physically depends
on the way space is curved by energy. For example, in Nordström's
theory of gravitation of 1913 \cite{Nordstrom1916}, the two previous
contributions precisely cancel each other and give a deviation of
light which is identically zero~\footnote{In a modern point of view,
  this is due to the fact that the Nördstrom theory is a scalar theory
  $\phi$, and that the coupling Lagrangian should be $\phi T$ with $T$
  the trace of the energy-momentum tensor. For an electromagnetic
  field, this trace is zero, and therefore ligh cannot be coupled to a
  scalar.}, in contradiction with the experiment of 1919
\cite{Dyson1919}. Nevertheless, Nordström's theory is theoretically
viable, fully relativistic and in accordance with the equivalence
principle.

However, one of the lessons of special relativity is that space and
time are profoundally linked into a spacetime concept. Therefore, it
seems natural to apply to space what has been observed with time~: if
duration depends on the gravitational field, length should also
depends on gravitational field. The question is to know what
modification should be done on length. Going back to equation (\ref
{eq: tau phi}), we can write the relation between $ \dd \tau $ and
$ \dd \tau_\Phi $ as a time dilation relation. Indeed, by posing
$ \Phi = -2 v_G^2 $, one has~:
\begin{equation}
  \label{eq: 15}
  \dd \tau_0 = \gamma_\Phi \dd \tau_ \Phi
\quad\hbox{with}\quad
\gamma_ {\Phi} = \left (1 + {2 \Phi \over c^2} \right)^{-1/2} =
\left (1 - {v_G^2 \over c^2} \right)^{- 1/2} \geq 1
\end{equation}
The duration in a distant observer is dilated. One can try a
contraction of length in the radial direction, that is to say in the
direction in which the gravitational potential varies. We would then
have~:
\begin{equation}
  \label{eq:1}
\dd r_0 = \frac {\dd r_ \Phi} {\gamma_ \Phi}  
\end{equation}
with $\dd r_\Phi$ the length travelled during time $\dd\tau_\Phi$ at
the level of the particle of light, while $\dd r_0$ is the length as
seen by a distant observer. Then, instead of starting from (\ref {eq:
  13}), we have to require, because of the equivalence principle,
\begin{equation}
  \label{eq: 14}
c = \frac {\dd r_ \Phi} {\dd \tau_ \Phi}
\end{equation}
So that the phase velocity would be given by
\begin{equation}
  \label{eq: 14bis}
c_ {p, \Phi} = \frac {\dd r_0} {\dd \tau_0} =
\frac {1} {\gamma_\Phi^2} \frac {\dd r_ \Phi} {\dd \tau_ \Phi} =
\frac {c} {\gamma_\Phi^2} = c \left (1 + \frac {2 \Phi} {c^2} \right)
\end{equation}

This is the new phase velocity measured by a distant observer. We
obtain the same relation as the equation~(\ref{eq: 12}), simply
replacing $ \Phi $ with $ 2 \Phi $. It is worth noting that the radial
contraction of equation (\ref{eq:1}) is nothing else that a space
curvature. This contraction also define the right direction of the
parallel transport of the photon~\cite{Ferraro2003}.

The previous result is retrieved, in a more modern way, by considering
the following modification of the square of the interval~:
 \begin{equation}
\dd s^2 = \left (1 - {r_S \over r} \right) \, c^2 \dd t^2 -
{1 \over 1- {r_S / r}} \, \dd r^2
\quad\hbox{with}\quad
{r_S \over r} = {2GM  \over r c^2} = - {2 \Phi \over c^2}
\end{equation}
We recover the space-time interval of the Schwarzchild metric proposed
by the latter in 1916 \cite{Schwarzchild}. The trajectory of the light can be
obtained according to $\dd s^2$ and therefore
\begin{equation}
c _ {\Phi} = {\dd r \over \dd t} = {c \over \gamma_ {\Phi}^2} =
c \left (1 + {2 \Phi \over c^2} \right) =
c \left (1 - {r_S \over r} \right)
\end{equation}
This expression of $ c_ {\Phi} $ looks like the one obtained initially
by Einstein in 1911, with the factor $ 2 $ which affects the
gravitational potential. It is then sufficient to use Einstein's wave
reasoning to obtain a double deviation angle, in accordance with the
observations \cite{Dyson1919}.

Let us notice that other choices were {\it a priori} admissible. For
example, in the Nordström theory, this choice would be not to contract
the radial lengths, but on the contrary to expand them,
$ \dd r_0 = \gamma_ {\Phi} \dd r_ {\Phi} $. This amount to treat space
and time with the same factor. In modern langage, it means that the
metric is conformally flat, that is to say
$\dd s^2=f(\Phi)(c^2\dd t^2-\dd r^2)$. This would give
$ c_ {p, \Phi} = c $. Therefore, in this theory, because the phase
velocity is a constant, light is not bended. Physically, there is a
perfect compensation between the effect on time and the effect on
space.

\section*{Conclusion}

Let's remember the two essential points.

{\it i)} From Newtonian perspective, Soldner showed as early as 1801
that light should be deflected by a spherical mass. This deviation is
identical (at lowest order) to the one obtained by Einstein in 1911,
although their approaches differ substantially. Equality of both
results comes from the principle of equivalence and the link between
the velocity of a physical object  and the velocity of the associated de
Broglie wave.

{\it ii)} An intuitive reasoning based on the effect of a
gravitational potential on radial lengths and thus on the curvature of
space could have put Einstein on the track of general relativity as
early as 1911; he would then have found the right result for light
deviation, and at the same time the Schwarzschild metric (see
also~\cite{Provost2016}).

\bigskip \bigskip


\newpage{\pagestyle{empty}\cleardoublepage}

\end{document}